\documentclass[%
 aip,
 amsmath,amssymb,
 reprint,%
]{revtex4-1}

\usepackage{graphicx}% Include figure files
\usepackage{bm}% bold math
\usepackage[mathlines]{lineno}% Enable numbering of text and display math
\usepackage{mathtools}
\DeclarePairedDelimiter\bra{\langle}{\rvert}
\DeclarePairedDelimiter\ket{\lvert}{\rangle}
\DeclarePairedDelimiterX\braket[2]{\langle}{\rangle}{#1 \delimsize\vert #2}

\usepackage[utf8]{inputenc}
\usepackage[T1]{fontenc}
\usepackage{color,soul}

\usepackage{mathptmx}
\usepackage{etoolbox}
\makeatletter
\def\@email#1#2{%
 \endgroup
 \patchcmd{\titleblock@produce}
  {\frontmatter@RRAPformat}
  {\frontmatter@RRAPformat{\produce@RRAP{*#1\href{mailto:#2}{#2}}}\frontmatter@RRAPformat}
  {}{}
}%
\makeatother
\usepackage{babel}

\DeclareMathOperator{\Var}{Var}

\begin{document}

\preprint{AIP/123-QED}

\title{Bayesian Estimation for Bell State Rotations}

\author{Luke Anastassiou}
 \email{Luke.Anastassiou@liverpool.ac.uk}
 \affiliation{Department of Electrical Engineering and Electronics, University of Liverpool,  Brownlow Hill, Liverpool, L69 3GJ, UK.}
\author{Jason F. Ralph}
 \email{jfralph@liverpool.ac.uk}
 \affiliation{Department of Electrical Engineering and Electronics, University of Liverpool,  Brownlow Hill, Liverpool, L69 3GJ, UK.}
\author{Simon Maskell}
 \email{smaskell@liverpool.ac.uk}
 \affiliation{Department of Electrical Engineering and Electronics, University of Liverpool,  Brownlow Hill, Liverpool, L69 3GJ, UK.}
\author{Pieter Kok}
 \email{p.kok@sheffield.ac.uk}
 \affiliation{Department of Physics, University of Sheffield, Hounsfield Rd, Sheffield S3 7RH, UK.}

\date{\today}

\begin{abstract}
This paper explores the effect of three-dimensional rotations on two-qubit Bell states and proposes a Bayesian method for the estimation of the parameters of the rotation. We use a particle filter to estimate the parameters of the rotation from a sequence of Bell state measurements and we demonstrate that the resultant improvement over the optimal single qubit case approaches the $\sqrt{2}$ factor that is consistent with the Heisenberg limit. We also demonstrate how the accuracy of the estimation method is a function of the purity of mixed states.
\end{abstract}

\maketitle

\section{Introduction}
\label{sec:sec1}
Metrology is one of the key areas where quantum physics could provide significant benefits over classical systems. Metrology underpins all sensing systems and it provides fundamental limits to the accuracy that the parameters representing physical quantities can be measured~\cite{Cramer1999}. Quantum metrology as a distinct discipline goes back many years, and Jonathan Dowling was one of its loudest and most engaging of advocates~\cite{Dowling2014}: ``{\em Quantum metrology has been found to enable measurements with a precision that surpass the classical limit, and has grown into an exciting new area of research with potential applications... in gravitational wave detection, quantum positioning and clock synchronization, quantum frequency standards, quantum sensing, quantum radar and LIDAR, quantum imaging and quantum lithography}''. 

The physical properties underpinning quantum metrology are the non-classical aspects of quantum states, such as entanglement. In its simplest form, classical estimation processes are limited by the `shot noise limit' (SNL), where the accuracy of measurements of parameters is limited by a factor that is proportional to $1/\sqrt{N}$, and where $N$ is the number of measurements (or `resources'). By comparison, a suitable quantum measurement process can provide an accuracy that is proportional to $1/N$, which is called the `Heisenberg limit' (HL). However, in practice, obtaining benefits using quantum metrology requires the ability to create complex entangled states, and to maintain them in the presence of environmental influences. The benefits of quantum metrology are limited by the level of entanglement present and decoherence due to the environment.

Here, we consider a specific example where an entangled quantum state can provide an accuracy benefit that approaches the Heisenberg limit, and we present a Bayesian estimation method that could realise this benefit. The example discussed is a theoretical model for a gyroscope, where the effect of a rotation is represented by the rotation of the system's measurement axes with respect to a prepared quantum state. Specifically, we consider two level systems (qubits) that are spin-$1/2$ angular momentum states. These systems can be prepared either as individual spin states or in pairs as Bell states. The period between preparation and measurement is small, so that the systems' angular momentum is preserved, and so that rotations and any mixing due to environmental effects are relatively small. We demonstrate that Bayesian estimation can be used to estimate the three-dimensional Euler rotation angles (heading/yaw, pitch, and roll) using a sequence of projective measurements, and that a particle filter approach to Bayesian estimation~\cite{Gordon1993,Arulampalam2002,Cappe2007} for a sequence of measurements on the entangled Bell pairs can provide an improvement in accuracy that approaches a factor of $\sqrt{2}$ when compared to single spin states. This is expected, in that Bell states only contain entanglement between two two-state systems.

Gyroscopes are a key element of modern inertial navigation systems~\cite{Titterton2004,Groves2013}. They measure angle rates, which are integrated to obtain the Euler angles, which are used in turn to resolve the specific forces measured by accelerometers to obtain estimates for the translation motion of a platform. These systems form the basis for all high bandwidth navigation systems used in aerospace, maritime and space applications~\cite{Peshekhonov2011,Ma2020}. However, the main focus of the current paper is not the practical realisation of a gyroscope. Rather, the aim is to explore the benefits of using an entangled state and to demonstrate a practical Bayesian approach to the estimation problem. 

Quantum sensing of rotations has been the subject of a lot of attention. With standard optical gyroscopes, such as ring laser gyroscopes and fibre optical gyroscopes~\cite{Titterton2004}, being used in modern navigation systems, it is natural to ask whether quantum states of light could be used to enhance their performance~\cite{Fink2019}. The Sagnac effect, which forms the basis for optical gyroscopes\footnote{Although it should be noted, that neither ring laser nor fibre optical gyroscopes operate using the commonly presented Sagnac effect~\cite{Culshaw1983,Nayak2011}}, can be posited as a phase measurement problem, where quantum states are known to show significant benefit in terms of accuracy. In addition, quantum sensors based on cold atom physics are currently being developed and can show remarkable sensitivity to rotations~\cite{Gustavson2000,Dutta2016}. Such systems currently use the Sagnac effect for matter waves, but further benefits could arise if more sophisticated quantum states of matter waves could be prepared~\cite{Scully1993,Dowling1998}.

Quantum estimation methods for rotations have been studied previously and Bayesian methods for two-level systems have been developed, for single two-level systems and clusters of qubits forming error correcting codes~\cite{Martinez2019}. In Ref~\cite{Martinez2019}, the problem was formulated as a phase estimation process, but the single qubit method can also be applied to the problem discussed in this paper where rotation generates a phase shift of the single qubit states. More complex quantum states have also been studied, including coherent and anti-coherent states~\cite{Goldberg2018,Goldberg2019,Goldberg2020a,Martin2020}, Majorana constellations and `Queens of Quantum'~\cite{Giraud2010,Bjork2015,Goldberg2020b,Goldberg2021}. Other approaches to quantum rotation sensing have adopted an approach that is motivated by quantum computation~\cite{Mo2019}.

The Bell pair states used in this paper are distinct from the Majorana and more complex states based on symmetries of angular momentum in that the restriction of the quantum states to pairs of two-level systems means that the Majorana constellation for spin-$1$ states is relatively uninteresting (formed from two antipodal points)~\cite{Giraud2010,Bjork2015,Goldberg2020b}. However, as we will show, the probability distributions for projective measurements of Bell states along rotated axes show some interesting properties which are useful for the estimation of three-dimensional rotations. 

The measurement of a physical rotation can be achieved in a number of ways depending on the application and implementation of the system. Within the context of spin-measurements, a quantum tomographic approach may be seen as a natural tool to achieve such an estimation. Any system that experiences a physical rotation in three-dimensional space may be described by rotations around the Bloch sphere under the condition that the Bloch sphere describing the quantum system in some way maps onto the physical rotations of the system, or at the very least, relates to the Bloch sphere rotations. Hence, all literature concerning quantum state estimation and quantum tomography becomes a relevant consideration. Over the years a variety of popular quantum tomographic techniques in the literature have been found to achieve different advantages and optimality criteria. In single qubit tomography, Ref~\cite{Sugiyama2012}, shows that through an adaptive measurement update method based on average variance optimality, an analytical solution was found for the estimation of a single qubit mixed state using projective measurements. Such a solution was demonstrated to result in a reduction of the complexity of measurement updates. In Ref~\cite{Huszar2012}, it was demonstrated that an adaptive optimal experimental scheme based on Bayesian inference could drastically reduce the total number of measurements required to estimate a single qubit quantum state. Different approaches to such a task offer different benefits, the former by reducing computational effort. By extending the discussion to entangled systems and starting with the simplest kind (Bell states), matters become more complex due to the nature of Bell measurements. Distinguishing between different Bell states, and taking account of the self-connected nature relating Bell states to each other presents both a challenge and an opportunity for exploration~\cite{Kim2001}. It may be agreeable that a stepping-stone to better understanding the nature of these fundamental units of correlation is to explore the nature of rotations affecting Bell states.

We start in section~\ref{sec:sec2} by first outlining the rotational properties of the Bell states for spin-$1/2$ systems. In section~\ref{sec:sec3}, following Ref~\cite{Martinez2019}, we outline the optimal Bayesian estimation process for single spin-$1/2$ states, and show how these generalise to the case of Bell states. We also discuss the appropriate ways to measure the resources used in the measurements of rotation in the single spin and Bell state cases. In section~\ref{sec:sec4}, we present the results for a pure state, and then consider the effect of mixed states in section~\ref{sec:sec5}, before summarising for key elements of the paper in the concluding section. 

\section{Rotations of Bell states}
\label{sec:sec2}
The model system considered in this paper is where a Bell state is constructed from spin-$1/2$ systems in a specified basis. We assume that the main effect of rotation is to reorient the measurement system and the axes in which any measurement is made. We also assume that any rotation is applied sufficiently rapidly that the angular momentum of the Bell state is constant, with respect to the original axes in which it has been prepared\footnote{This is the basis for the mechanical gyroscopes used in older gimballed navigation systems, which use the conservation of angular momentum to maintain a static reference frame for rotations.}, and that any rotations are relatively small. The constraint on the size of the rotation is not strictly necessary, but it does simplify the presentation of the estimation method presented here.  Later, we will find that the optimal measurements of Bell states are not necessarily along the axes in which they are prepared so we will start by discussing the properties of Bell states constructed from spin-$1/2$ systems defined in arbitrary axes and their properties under the action of rotation.

The simplest form of entanglement is when two systems, $A$ and $B$ each with two levels are entangled together, forming a Bell state\cite{}. Four different Bell states exist, and together they form the Bell basis. Below we study the properties of their rotations, either when both subsystems undergo the same rotation, or when one subsystem undergoes a rotation. The four Bell states are given by:
\begin{eqnarray}\label{BellStates}
\ket{\Psi^\pm}&=\frac{1}{\sqrt{2}} \left( \ket{01} \pm \ket{10} \right),&\\
\ket{\Phi^\pm}&=\frac{1}{\sqrt{2}} \left( \ket{00} \pm \ket{11} \right).&
\end{eqnarray}
When expressed in this way, it is usually implied that the Bell states are in the $z$-basis, which is the conventional quantization axis. A single qubit may be rotated by applying a rotation operator of the form \cite{Audretsch2007}:
\begin{equation}
\bm{R}_j(\theta) = e^{-i\theta\bm{\sigma}_j /2}, \hspace{0.5cm} j = x, y, z,
\end{equation} 
where $\theta$ is the angle rotated about the $x, y$ or $z$ axis, and $\sigma_j$ is the Pauli operator used to perform the rotation. Since a Bell state is composed of two systems A and B, we can use a rotation operator of this form to construct a rotation operator $\bm{\mathcal{R}}_j(\theta)$ which will perform a rotation on either one or both subsystems. The form of this operator will vary depending on what we wish to achieve. For rotations of just one subsystem, our operator can be defined as,
\begin{equation}
    \bm{\mathcal{R}}_j(\theta) := \bm{R}_j(\theta)^A \otimes \bm{I}^B, \hspace{0.2cm} \text{or} \hspace{0.2cm} \bm{\mathcal{R}}_j(\theta) := \bm{I}^A \otimes \bm{R}_j(\theta)^B,
\end{equation}
depending on whether system A or B is rotating. Additionally, we can define an operator which rotates the entire Bell state equally:
\begin{eqnarray}
\bm{\mathcal{R}}_j(\theta) & := &\bm{R}_j^A \otimes \bm{R}_j^B  \nonumber \\ 
& = &\left( \cos\left( \tfrac{\theta}{2}\right)\bm{I} - i\sin\left( \tfrac{\theta}{2}\right) \bm{\sigma}_j^A \right)\nonumber\\ 
&& \otimes \left( \cos\left( \tfrac{\theta}{2}\right)\bm{I} - i\sin\left( \tfrac{\theta}{2}\right) \bm{\sigma}_j^B \right) \nonumber\\
& = &\cos^2 \left( \tfrac{\theta}{2} \right) \bm{I} \otimes \bm{I} \nonumber\\ 
&& - i\cos\left(\tfrac{\theta}{2}\right)\sin\left(\tfrac{\theta}{2}\right) \left[ \bm{I}\otimes\bm{\sigma}_j^B + \bm{\sigma}_j^A \otimes \bm{I} \right]\nonumber\\ 
&& - \sin^2 \left(\ \tfrac{\theta}{2} \right) \bm{\sigma}_j^A \otimes \bm{\sigma}_j^B,  \label{rotop:1}
\end{eqnarray}
where we have used the identity:
\begin{equation}
e^{-i\theta\bm{\sigma}_j /2} = \cos(\tfrac{\theta}{2})\bm{I} - i \sin(\tfrac{\theta}{2}) \bm{\sigma}_j,
\end{equation}
which is valid for an operator of this kind\cite{}. The rotation operator in \eqref{rotop:1} describes a rotation affecting each subsystem equally, hence it will be the most relevant for our considerations in the study of the effect of three-dimensional rotations on our Bell states. Substituting $x$, $y$, and $z$, we can study the result of rotating each Bell state about the different axes, shown in Table~\ref{tab:bellrots}.
\begin{table}[h]
\begin{tabular}{|c|| c | c | c | c |}
\hline
j & $\hat{\mathcal{R}_j} \ket{\Psi^-}$ & $\hat{\mathcal{R}_j} \ket{\Psi^+}$ & $\hat{\mathcal{R}_j} \ket{\Phi^+}$ & $\hat{\mathcal{R}_j} \ket{\Phi^-}$ \\ \hline\hline
$x$ & $\ket{\Psi^-}$ & 
$\begin{array}{l}\cos\theta\ket{\Psi^+} \\ -i\sin\theta\ket{\Phi^+}\end{array}$ & $\begin{array}{l}\cos\theta\ket{\Phi^+} \\ -i\sin\theta\ket{\Psi^+}\end{array}$ & $\ket{\Phi^-}$ \\ \hline
$y$ & $\ket{\Psi^-}$ & 
$\begin{array}{l}\cos\theta\ket{\Psi^+} \\-\sin\theta\ket{\Phi^-}\end{array}$ & $ \ket{\Phi^+} $ & $\begin{array}{l}\cos\theta\ket{\Phi^-} \\+\sin\theta\ket{\Psi^+}\end{array}$ \\ \hline
$z$ & $\ket{\Psi^-}$ & $\ket{\Psi^+}$ & 
$\begin{array}{l}\cos\theta\ket{\Phi^+} \\ -i\sin\theta\ket{\Phi^-}\end{array}$ & 
$\begin{array}{l}\cos\theta\ket{\Phi^-} \\ -i\sin\theta \ket{\Phi^+}\end{array}$ \\ \hline
\end{tabular}
\caption{The result of applying a rotation operator $\hat{\mathcal{R}}_j(\theta) := \bm{R}_j^A(\theta) \otimes \bm{R}_j^B(\theta)$,  $j = x,y,z$ to the four Bell states. }
\label{tab:bellrots}
\end{table}
The rotations indicate that some Bell states are invariant under rotations about certain axes, while the $\ket{\Psi^-}$ Bell state is invariant under all rotations. The former suggests a similarity in the mathematical structure of the relationship between the Bell state rotations and an ordinary single system rotation, where we find a directional preference when it comes to state rotations about particular axes. Further to this, we notice that for the non-invariant rotations, axial rotations of this kind transform one Bell state into another.
More generally, if we wish to describe a rotation $\bm{\theta}$ of the Bell state about some unit vector $\bm{k}$ describing the axis of rotation ($\bm{\theta} = \theta \bm{k} = (\theta_x, \theta_y, \theta_z)$), we may use the rotation operator given by:
\begin{equation}
\bm{\mathcal{R}}(\bm{\theta}) := \bm{R}(\bm{\theta}) \otimes \bm{R}(\bm{\theta}) = e^{-i\theta \bm{k}\cdot \bm{\sigma} /2} \otimes e^{-i\theta \bm{k}\cdot \bm{\sigma} /2},
\end{equation}
where $\bm{\sigma}$ is the Pauli vector. Applying the identity:
\begin{equation}
e^{-i\theta\, (\bm{k}\cdot \bm{\sigma}) /2} = \cos(\tfrac{\theta}{2})\bm{I} - i \sin(\tfrac{\theta}{2}) \,(\bm{k}\cdot\bm{\sigma}),
\end{equation}
we obtain the following rotation formulae:
\begin{eqnarray}
\bm{\mathcal{R}}(\theta)\ket{\Phi^+} &=& \left( \cos\theta - k_y^2 \cos\theta + k_y^2 \right) \ket{\Phi^+} \nonumber\\
&& +\left( -ik_x\sin\theta -ik_yk_z +ik_yk_z\cos\theta \right) \ket{\Psi^+} \nonumber\\
&& +\left( -ik_z\sin\theta + ik_xk_y - ik_xk_y\cos\theta \right) \ket{\Phi^-}, \label{eq:RPhip}\\
\bm{\mathcal{R}}(\theta)\ket{\Phi^-} &=& \left( \cos\theta -k_x^2\cos\theta +k_x^2 \right) \ket{\Phi^-} \nonumber\\
&& + \left( k_y\sin\theta -k_xk_z +k_xk_z\cos\theta \right) \ket{\Psi^+} \nonumber\\
&& + \left( -ik_z\sin\theta -ik_xk_y +ik_xk_y\cos\theta \right) \ket{\Phi^+}, \label{eq:RPhim}\\
\bm{\mathcal{R}}(\theta)\ket{\Psi^+} &=& \left( \cos\theta - k_z^2 \cos\theta + k_z^2 \right)\ket{\Psi^+} \nonumber\\
&& +\left( -ik_x\sin\theta +ik_y k_z -ik_yk_z \cos\theta \right) \ket{\Phi^+} \nonumber\\
&& +\left( -k_y\sin\theta - k_xk_z + k_xk_z\cos\theta \right) \ket{\Phi^-},\label{eq:RPsip}\\
\bm{\mathcal{R}}(\theta)\ket{\Psi^-} &=& \ket{\Psi^-}.
\end{eqnarray}
As expected for all pure states, the four Bell states remain maximally entangled after a rotation about any chosen axis, confirmed by computing the von Neumann entropy $S(\rho)$, equal to $1$ for all four Bell states.

We now consider a case where we construct a Bell state using states from the $z$-basis and then perform a measurement that distinguishes between different Bell states that can be constructed using states from a different basis. We can define an axis using an arbitrary unit vector lying on a sphere, and construct Bell states using this basis. Constructing an initial Bell state using the $z$-basis, we can perform a Bell measurement using a basis defined by two angles, the elevation ($\Theta = 0\ldots \pi$) and the azimuth ($\Lambda = 0\ldots 2\pi$). 

\begin{figure}
    \centering
    \includegraphics[width=1.0\hsize]{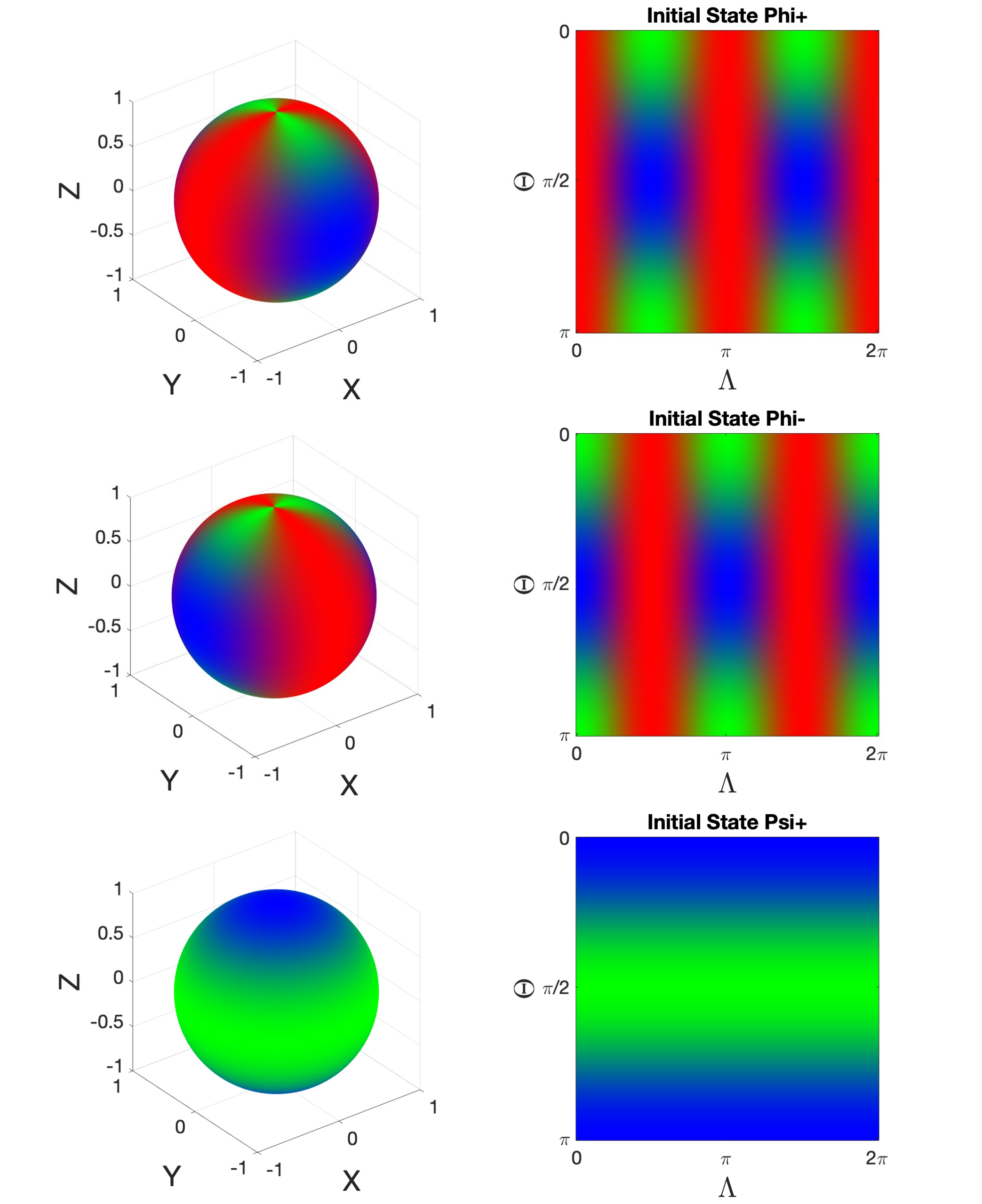}
    \caption{Regions showing the probability of each result from a Bell measurement on initial states -- $\ket{\Phi^+}$ (top row), $\ket{\Phi^-}$ (middle row), and $\ket{\Psi^+}$ (bottom row) -- constructed using the z-basis, where the Bell measurement uses a basis aligned along an axis defined by a vector on the sphere (left) or in terms of elevation-azimuth ($\Theta-\Lambda$). The colours indicate the probabilities to obtain each of the three Bell States, $\ket{\Phi^+}$ (red), $\ket{\Phi^-}$ (green), and $\ket{\Psi^+}$ (blue), from a measurement as a function of the measurement basis.}  
    \label{fig:spheres}
\end{figure}

\begin{figure}
    \centering
    \includegraphics[width=0.9\hsize]{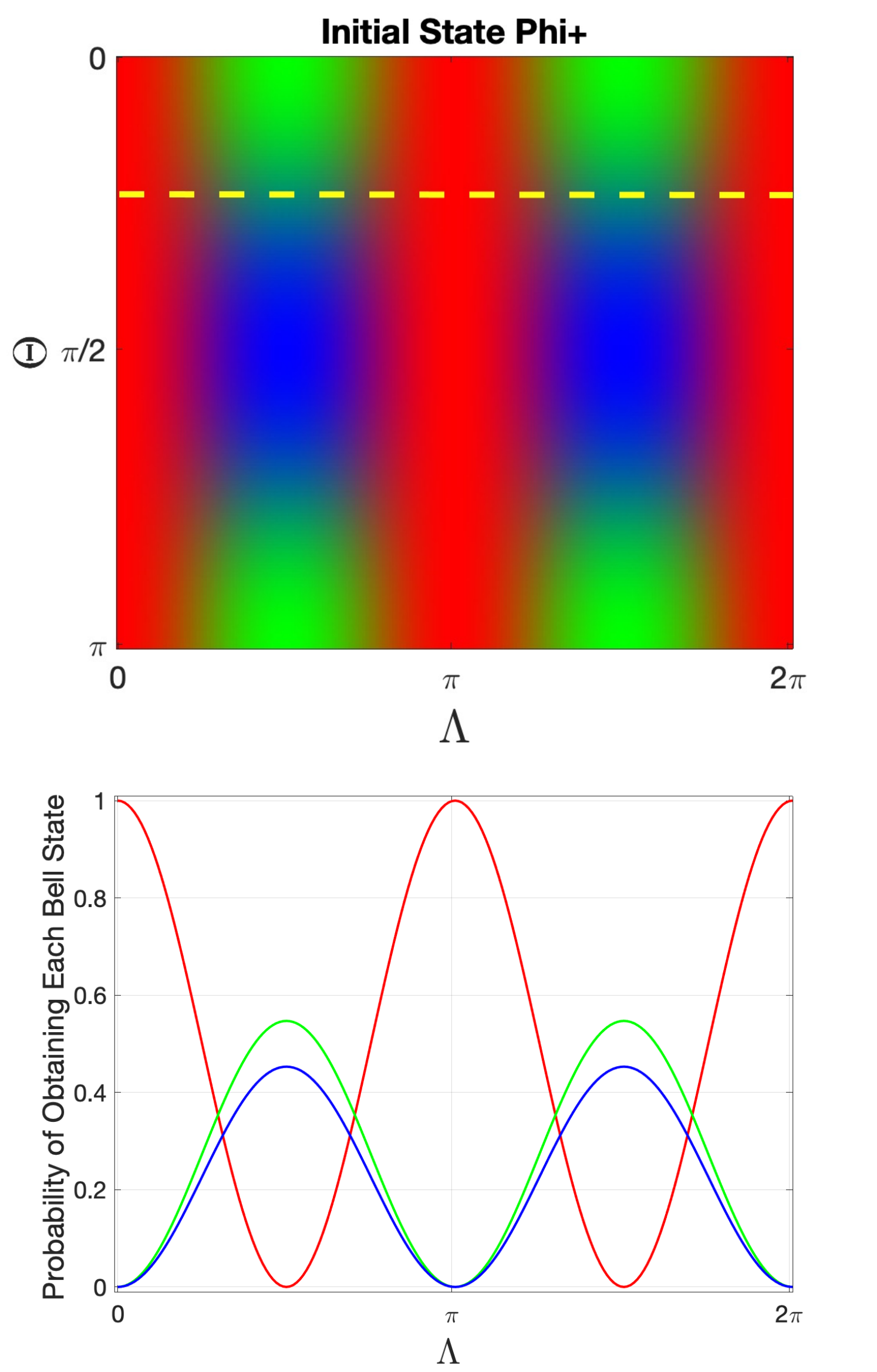}
    \caption{Example showing the probabilities to obtain each of the three Bell states -- $\ket{\Phi^+}$ (red), $\ket{\Phi^-}$ (green), and $\ket{\Psi^+}$ (blue) -- for a Bell measurement on an initial $\ket{\Phi^+}$ state as a function of the measurement basis (top), and as a function of $\Lambda$ for the value of $\Theta$ indicated by the yellow dashed line (bottom).}  
    \label{fig:probabilities}
\end{figure}

Figure~\ref{fig:spheres} uses colours to indicate the probability for obtaining one of the three Bell states ($\ket{\Phi^{\pm}}$ and $\ket{\Psi^+}$) for the three initial ($z$-basis) states affected by rotations ($\ket{\Phi^{\pm}}$ and $\ket{\Psi^+}$). The intensity of each Red-Green-Blue (RGB) colour indicates the probability for the corresponding Bell state to be obtained -- a pure colour is a probability equal to one or close to one, and darker/mixed tones correspond to lower probabilities shared between multiple possible results (see Figure~\ref{fig:probabilities} for an example of the probability values). We do not show the results for an initial $\ket{\Psi^-}$ state because the result would always be another $\ket{\Psi^-}$ state, since it is not affected by rotation, and none of the three other Bell states can result in the measurement of a $\ket{\Psi^-}$ state. 

\begin{figure}
    \centering
    \includegraphics[width=1.0\hsize]{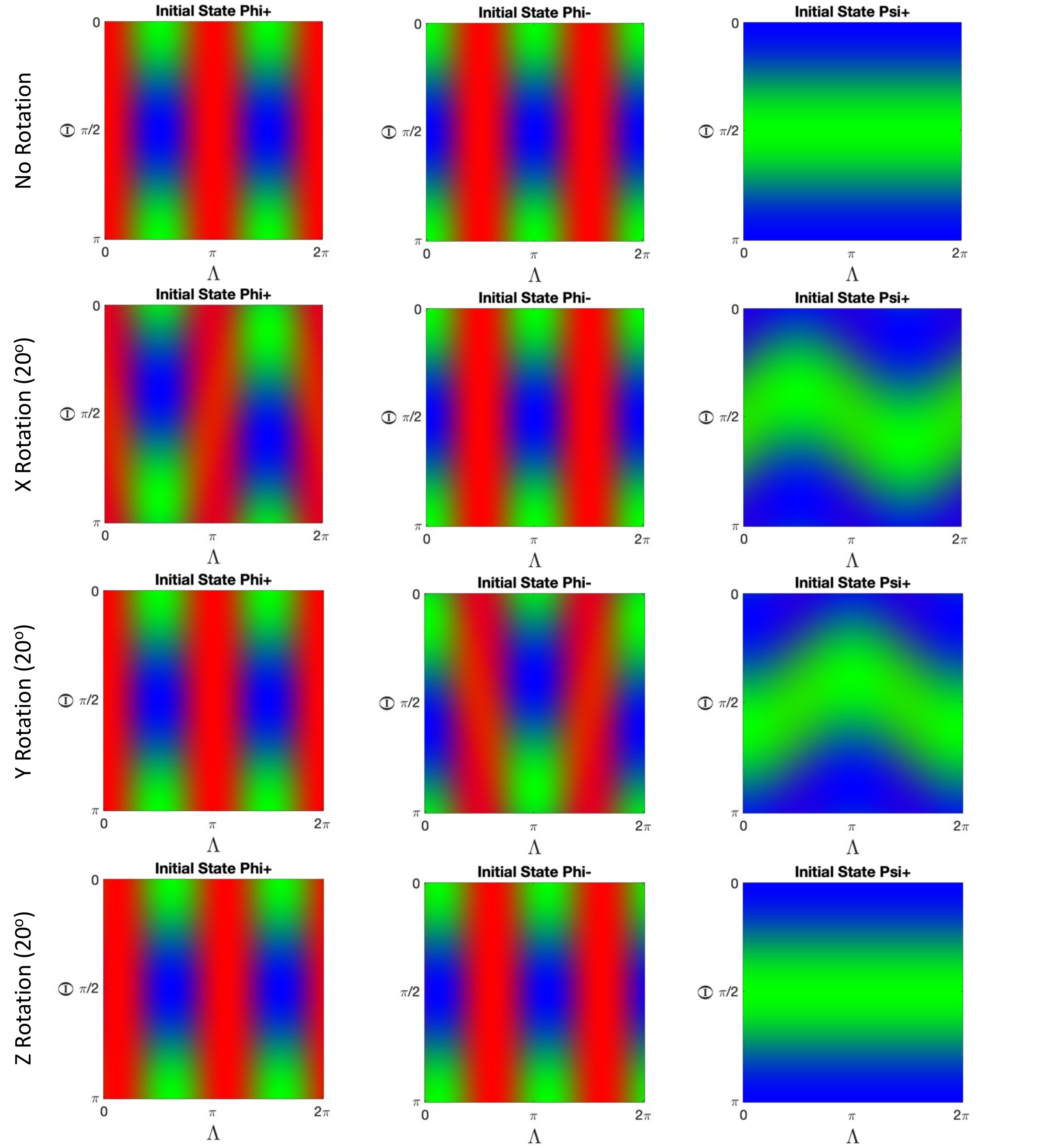}
    \caption{Matrix of colour plots showing the effect of individual rotations about each axis on the probability distributions for the initial Bell states: $\ket{\Phi^+}$ (left column), $\ket{\Phi^-}$ (centre column), and $\ket{\Psi^+}$ (right column). The colours indicate the probabilities to obtain each of the three Bell States, $\ket{\Phi^+}$ (red), $\ket{\Phi^-}$ (green), and $\ket{\Psi^+}$ (blue), from a measurement as a function of the measurement basis.}  
    \label{fig:tableofrotations}
\end{figure}

From Figure~\ref{fig:spheres}, we can see that an initial $\ket{\Phi^{\pm}}$ state can lead to any of the three Bell States $\ket{\Phi^{\pm}}$ or a $\ket{\Psi^+}$ after measurement, but a $\ket{\Psi^+}$ state cannot lead to a $\ket{\Phi^+}$. Figure~\ref{fig:probabilities} shows an example of the probabilities of obtaining each of the Bell states given an initial $\ket{\Phi^+}$ state as a function of the orientation of the measurement basis, and a plot of these probabilities along a line at constant $\Theta$ and varying $\Lambda$. 

In fact, the two $\Phi$ states have very similar patterns of probabilities, although the $\ket{\Phi^+}$ pattern is rotated $\pi/2$ around the z-axis relative to the $\ket{\Phi^+}$ case. We also note that there are specific measurement axes on these patterns where each of the three Bell states, $\ket{\Phi^{\pm}}$ and $\ket{\Psi^+}$, are equally likely to be obtained from a measurement. These points of equal probability offer the potential to maximise the information extracted from each measurement, since the information from each measurement is closely related to the number and the probability of each of the possible outcomes. 

Now we turn to the effect of rotation on the distributions for Bell measurements on the spheres shown in Figure~\ref{fig:spheres}. For small rotations, characterised by a rotation $\bm{\theta} = (\theta_x, \theta_y, \theta_z)$, we can examine the effect on the probability distributions of rotations about each axis in turn. Figure~\ref{fig:tableofrotations} shows these effects on the probability distributions to obtain the different Bell states for rotations of $20$ degrees about each axis -- we show $20$ degree rotations because the changes for much smaller rotations are very subtle and would be difficult to see in the Figure. 

In Figure~\ref{fig:tableofrotations}, we can see that an initial $\ket{\Phi^+}$ state (first column) is sensitive to rotations about the $x$-axis and the $z$-axis, but not the $y$-axis. The $\ket{\Phi^-}$ state (second column) is sensitive to rotations about the $y$-axis and the $z$-axis, but not the $x$-axis. And the $\ket{\Psi^+}$ state (third column) is only weakly sensitive to rotations about the $x$-axis and the $y$-axis, but not the $z$-axis. 

Here, we will focus on small rotations, typically less than about $10$ degrees or about $0.1745$ radians, and use the points on the spheres where the three Bell states are equally, or nearly equally, likely. Using a similar approach for larger rotations would be possible, but the approach would become more complicated. For example, as the rotations about the $x$-axis and $y$-axis approach $45$ degrees, the $\ket{\Psi^+}$ initial state becomes more sensitive to rotations and as it approaches a $90$ degrees of rotation, the pattern for the $\ket{\Psi^+}$ state swaps with one of the $\ket{\Phi}$ states: for example, at $90$ degrees of rotation around the $x$-axis the probability distributions for the $\ket{\Psi^+}$ and the $\ket{\Phi^+}$ interchange. To avoid these complications, we focus on small rotations so that each component has a standard deviation of less than about $\sigma_{\theta}=5$ degrees or approximately $0.873$ radians.

\section{Bayesian estimation of rotations}
\label{sec:sec3}
The aim of Bayesian estimation is to provide values for the parameters being estimated and an approximation for the probability distribution function (pdf) associated with these estimates (the posterior distribution, conditioned on the measurements that have been obtained). In the case of the well-known Kalman filter, which is used in many practical engineering systems, the (classical) state of the system is estimated alongside the estimated covariance for the errors in the state~\cite{Arulampalam2002,Cappe2007}. For linear systems and for systems that are not very nonlinear, the Kalman filter or adapted versions of the Kalman filter are often a very good and very efficient method for estimating parameters. With strongly nonlinear systems, other approaches based on particle-based approximations are often used~\cite{Arulampalam2002,Cappe2007}. Particle filters use a set of candidate solutions that represent possible values that the parameters could take; the `particles' represent points in the classical state/parameter space that evolve according to the same model as the underlying system. Each member of the set of possible solutions has a weight, which is updated after each measurement to reflect how likely it was to have been the true parameter value, given the value of the measurement. The less likely it is that the particle could have generated that particular measurement, the lower the weight associated with that particle will be and, over a sequence of measurements, any particles that conflict with the measured values will end up with very low weights. To avoid a few particles dominating the distribution (somewhat confusingly called `degeneracy'), it is necessary to resample the particles once the weight becomes concentrated on a subset of the particles. Resampling selects new candidate solutions preferentially from regions of the parameter space that are `close to' the high weight particles, but it does not preclude the selection of particles in regions of low weight. As measurements are obtained, the distribution of particles and weights should provide a robust estimate of the parameters and the associated pdf.

In this paper, we consider measurements on the quantum states to be projective measurements. A measurement sequence starts with a known pure state, which is subject to a rotation defined by the three Euler angles: heading ($\theta_x$), pitch ($\theta_y$) and roll ($\theta_z$). The quantum state is then measured, producing a measurement result and projecting the state onto the eigenstate associated with the measured value. Estimation of the three rotation parameters is done over a sequence of measurements using a particle filter, where each particle represents a possible value for the three Euler angles. The weights are updated based on the probability of getting each of the measured values, given the initial state and the rotation parameters associated with each particle. 

\subsection{Bayesian estimation with single spins}
The case of estimating rotations from measurements of single spin-1/2 states is relatively straightforward, and has been dealt with in Ref~\cite{Martinez2019}; where the probability distribution for a single rotation angle is estimated using an iterative method and a Gaussian assumption. In this case, to find one of the rotation angles, $\theta_y$ say, one initialises the state in an eigenstate of $\sigma_z$, the system undergoes a rotation, and one measures the $\sigma_x$ component of spin. The optimum method for generating all three components of the rotation vector is to cycle through a sequence for each of the axes: preparing a $z$ spin state, and measuring $x$ spin to estimate $y$ rotation; preparing $x$ spin state, and measure $y$ spin to estimate $z$ rotation; etc.. Cycling through the combinations in this way means that the state after the projective measurement will be in the desired basis for the next sequence of operations. 

For each of the components of $\bm{\theta}$, we have an independent sequence of measurements (three measurements per sequence) and an independent estimate. For component $\theta_y$, the measurements of $\sigma_x$ will either provide a $+$1/2 or a $-$1/2 eigenvalue, for $N_x$ measurements of $\sigma_x$ there will be $n_x^{(+)}$ measurements of $+$1/2 that can be represented as Bernoulli trial, with a probability of `successfully' obtaining a $+$1/2 eigenstate for a rotation angle of $\theta_y$ is given by,
\begin{equation}
p_x^{(+)} = \frac{1}{2}(1+\sin(\theta_y))
\end{equation}
and the estimate for the angle $\tilde{\theta}_y$ is given by,
\begin{equation}
\tilde{\theta}_y = \arcsin\left(\frac{2 n_x^{(+)}}{N_x}-1\right)
\end{equation}
This estimate provides an estimated probability of a success as
\begin{equation}
\tilde{p}_x^{(+)} = \frac{1}{2}(1+\sin(\tilde{\theta}_y))
\end{equation}
and estimated variances,
\begin{equation}
\Var(\tilde{p}_x^{(+)}) = \frac{\tilde{p}_x^{(+)}(1-\tilde{p}_x^{(+)})}{N_x}
\end{equation}
and 
\begin{equation}
\Var(\tilde{\theta}_y) = \frac{4 \tilde{p}_x^{(+)}(1-\tilde{p}_x^{(+)})}{N_x (1-(2\tilde{p}_x^{(+)}-1)^2)}= \frac{1}{N_x}
\end{equation}
For large numbers of measurements, where $\tilde{p}_x^{(+)}\rightarrow p_x^{(+)}$, the accuracy of the estimates of the angle will be proportionate to $1/\sqrt{N_x}$. 

\subsection{Estimation with Bell states}\label{estimation}
In comparison to the single spin case, the situation for Bell states is more complicated. If one looks at what the single spin case involves, one can see two key features. The preparation and the measurement axes are both orthogonal to the axis of rotation, and -- in the case that there is no rotation -- the probability of a `success' in the Bernoulli trial (obtaining the $+$1/2 eigenstate) is one half in each of the combinations in the sequences. It is possible to develop an approach for higher spin states similar to that described for one spin-1/2 state, with the preparation of eigenstates and orthogonal measurements. However, the information extracted from a measurement is generally maximised by having the probability shared equally among all of the available options, so we will explore the case with measurements of Bell states where the results are equally likely, or close to equally likely.

The probabilities of obtaining specific Bell states when making a Bell state measurement along a particular axis are shown in Figures~\ref{fig:spheres} and \ref{fig:tableofrotations}. Ignoring the $\ket{\Psi^-}$ case, which is insensitive to rotations, there are three possible states that could be distinguished by a Bell state measurement, but only $\ket{\Phi^+}$ or a $\ket{\Phi^-}$ will give a result that could result in any one of the three possible states\footnote{In fact, for larger rotations than those considered here, the three states can swap role, with the $\ket{\Psi^+}$ state swapping roles with one of the $\ket{\Phi}$ states. However, the distortions to the patterns of colours/probabilities are non-trivial and we have restricted consideration to small rotations to simplify the analysis presented in this paper.}. $\ket{\Psi^+}$ can only provide a $\ket{\Phi^-}$ or a $\ket{\Psi^+}$ state in a measurement. Looking at the distributions indicated in Figures~\ref{fig:spheres} and \ref{fig:tableofrotations}, starting in a $\ket{\Phi^{\pm}}$ state in the $z$-basis, there are specific points on the sphere where the probabilities are equal for obtaining all three of the relevant Bell states (where the three colours meet). Taking into account the fact that the axis goes through the sphere, there are four axes on each of the spheres for $\ket{\Phi^+}$ or a $\ket{\Phi^-}$ where the probabilities are equal.

In addition we note that, looking at Table~\ref{fig:tableofrotations} and Figure~\ref{fig:tableofrotations}, the two Bell states $\ket{\Phi^+}$ or a $\ket{\Phi^-}$ are sensitive to rotations around different axes. Both states are sensitive to rotations around the $z$-axis (the patterns in the top row of Figure~\ref{fig:tableofrotations} shift to the right), but $\ket{\Phi^+}$ is sensitive to rotations around the $x$-axis but not the $y$-axis (where the blue regions move up on the left and down on the right). By contrast, the $\ket{\Phi^-}$ state is sensitive to rotations around the $y$-axis but not the $x$-axis. This means that to characterise all three components of a rotation requires both $\ket{\Phi^+}$ and $\ket{\Phi^-}$ measurements, and to fulfil the requirement of equal probabilities of obtaining each of the three states, we select axes for the measurements to be as close as possible to the corners where the three colours meet. It is notable that these axes are not aligned to the basis in which the original Bell states are prepared (the $z$-basis in the cases presented here).

We can calculate the equal probability points numerically, and we only need to pick two of these, one for each of the initial states $\ket{\Phi^+}$ or a $\ket{\Phi^-}$. For $\ket{\Phi^+}$, we select the point $(\Theta_{\Phi^+},\Lambda_{\Phi^+}) = 0.95531662, 0.78539816)$ radians (corner point in the top left of the first Figure~\ref{fig:spheres} image), and we select the point $(\Theta_{\Phi^-},\Lambda_{\Phi^-}) = 0.61547971, 0.78539816)$ radians (top left of the second Figure~\ref{fig:spheres} image) for the $\ket{\Phi^-}$ initial state. Measurements are made alternating between an initial $\ket{\Phi^+}$ state (in the $z$-basis) state and an initial $\ket{\Phi^-}$ ($z$-basis) state, and Bell measurements along the axes defined by $(\Theta_{\Phi^+},\Lambda_{\Phi^+})$ and $(\Theta_{\Phi^-},\Lambda_{\Phi^-})$ respectively. 

To generate estimates for the three rotation components, we select an initial set of candidate rotations (particles in our particle filter, $\bm{\theta}_0^{(i)} = (\theta_{x,0}^{(i)}, \theta_{y,0}^{(i)}, \theta_{z,0}^{(i)})$, for $i = 1\ldots N_{\theta}$). For the small rotations considered here, the components for the initial particles are selected from independent Gaussian distributions with a standard deviation of $\sigma_{\theta} = 0.1745$ radians ($=10$ degrees), and they are all allocated an equal weight $w_0^{(i)}=1/N_{\theta}$. After each measurement, $m = 1\ldots N_{meas}$, $\ket{Z_m} \in \{\ket{\Phi^+}, \ket{\Phi^-}, \ket{\Psi^+}\}$, the weights are updated according to 
\begin{equation}
\tilde{w}_{m}^{(i)}=P\left(\ket{Z_m}\left|\right.\ket{\Phi^\pm}\right)w_{m-1}^{(i)}
\end{equation}
where $P(\ket{Z}|\ket{Z'})$ is the probability of obtaining a state $\ket{Z}$ from an initial $\ket{Z'}$ state. In practice, and to simplify the assessment of the behaviour of the estimation process when the states are mixed, we calculate the probabilities from the density matrices $P(\ket{Z}|\ket{Z'})=Tr(\rho_{Z}\rho_{Z'})$, where $\rho_{Z} = \ket{Z}\bra{Z}$. After all particles have been reweighted, the unnormalised weights ($\tilde{w}_{m}^{(i)}$) are renormalised so that they sum to one, $w_{m}^{(i)}=\tilde{w}_{m}^{(i)}/\sum_i \tilde{w}_{m}^{(i)}$. 

To avoid the weights accumulating on a small number of particles, we resample the particles whenever the effective number of particles $N_{\rm eff} = 1/(\sum_i (w^{(i)})^2)$ falls below a threshold value~\cite{Arulampalam2002,Cappe2007}, $N_{\rm eff} < N_{\theta}/2$. The particles are sampled from a distribution generated from the current particle weights -- a uniform random number between zero and one is used to select a particle by comparing the value to the cumulative weight distribution of the current particle weights~\cite{Arulampalam2002,Cappe2007}. To reduce the risk of degeneracy, we add a small perturbation to the selected particles by adding a random value selected from a Gaussian distribution, where the covariance of the Gaussian is related the current particle covariance, $\Sigma_m$. We use a defensive sampling approach~\cite{Green2017}, where the covariance is selected to be $0.1\Sigma_m$ in $90\%$ of resampled particles and $\Sigma_m$ in the remaining $10\%$ of cases. This ensures that the resampling adequately explores from the region of the parameter space around the current particle mean location. In addition, we perturb the particles slightly between measurements by adding process noise. Process noise is often used in filtering problems to perturb the solution to stop the solution (the estimate) becoming `stuck' in the wrong region of the solution space. We have found, empirically, that the near optimal form for the process noise in this system is Gaussian noise source with a gradually reducing standard deviation, $\sigma_m=0.1 m^{-\frac{2}{3}}$.

As the measurements are taken, the weights are updated and periodically resampled, and the particles gradually shift towards the true values of the rotations. The distribution of the particles and their weights provide a mean (estimated) value and an estimated error, calculated from the distribution of the particles and weights around the mean.

\subsection{Resource counting}
In order to assess the performance of an estimation procedure, we need to compare its mean square error to the amount of resources we have used. Using a lot of resources in a poor estimation procedure can yield a better mean square error than using only a few resources in a very good estimation protocol. We therefore need a careful resource count that allows us to compare different protocols meaningfully. The complication is that two estimation procedures may be using completely different physical systems, and it may not be obvious why one way to count the resources is better than another.

One method of universal resource counting for metrology was presented by Zwierz \emph{et al.}\cite{Zwierz2010,Zwierz2012}, and is based on a query complexity argument: how many times does the system sample the signal we wish to measure. If the signal causes a unitary transformation of the probe state, the expectation value of the interaction Hamiltonian $H$ of shifts in the parameter $\theta$ is a physical quantity that maps directly to the resources present in the probe system, provided we fix the energy scale such that the minimum energy eigenvalue is zero. For example, in order to find the resources present in a spin system to measure a simple rotation around the $z$ axis, we consider the unitary transformation of the rotation $U(\theta) = \exp(-i\theta S_z)$, with $S_z$ the generator of rotations around the $z$ axis for the spin state. For any spin state $\ket{\psi}$, the resources in the spin system that are used in the measurement of the rotation around the $z$ axis is $\bra{\psi} (S_z+\hbar s \mathbb{I})\ket{\psi}$. For a spin state in the maximum eigenvalue state $\ket{s,s}$ in the $z$-direction, the amount of resources used in the measurement is $2\hbar s$, which scales linearly with the spin $s$.

How can we compare the resources present in a single large spin system $s$ with the resources present in two smaller systems $s_1$ and $s_2$? Assuming that both spins undergo the same rotation, the interaction Hamiltonian will have the form $S^{(1)}_z \otimes\mathbb{I} + \mathbb{I}\otimes S_z^{(2)}$. Fixing the energy zero point for these spins, we find that the resource count for the separable state $\ket{s_1,s_1}\otimes\ket{s_2,s_2}$ is given by $2(s_1+s_2)\hbar$. This allows us to meaningfully compare the resources for rotation measurements using different physical systems.

In our case, we calculate the resources ($N_{\rm res}$) for a particular two-spin-$1/2$ state $\ket{Z}$ by summing over the traces of the density operator for that state with the generators of the spin states. That is,
\begin{equation}
    N_{\rm res} = \sum_j Tr(\rho_{Z} \sigma_j \otimes \sigma_j) \hspace{0.5cm} j = x, y, z\, .
\end{equation}
For a one-qubit state, the trace would only include a single Pauli operator. 

\section{Results}
\label{sec:sec4}
\begin{figure}
    \centering
    \includegraphics[width=0.95\hsize]{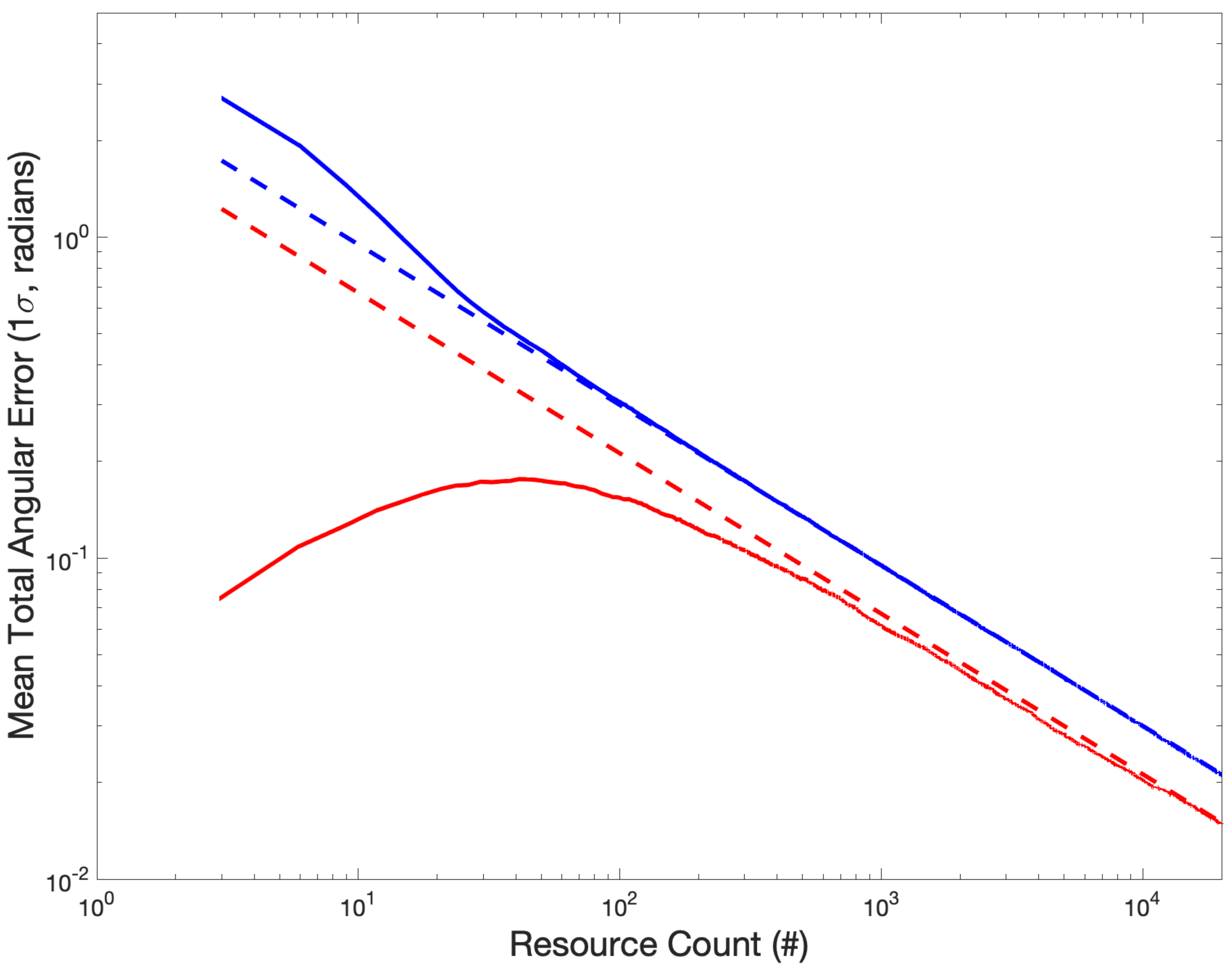}
    \caption{Mean Total Angular Errors (sum of errors for rotations around all three axes, 1 standard deviation) as a function of the resource count for single qubit measurements (blue) and Bell state measurements (red), with $1/\sqrt{N}$ (blue dash) and $1/\sqrt{2N}$ (red dash) shown for comparison.}
    \label{fig:errors}
\end{figure}

We have used the particle filter estimation method described in subsection~\ref{estimation} to calculate the total errors in the three rotational parameters (the Euler angle errors) as a function of the resource count. For small errors, as are considered here, the total errors are simply the root sum square of the individual components, which is then averaged over a thousand individual realisations/runs of the particle filter. The results are shown in Figure~\ref{fig:errors}, together with the results for the single qubit case and dashed lines to indicate the limiting values $1/\sqrt{N}$ (single qubit states) and $1/\sqrt{2N}$ (Bell states). 

The mean result for the Bell states are initially low and rises slightly as the number of resources increases. This reflects the use of a prior distribution, which is used to initialise the particle filter and reflects the fact that we have restricted consideration to small rotations. The errors grow as the particles are perturbed by the initial fluctuations caused by the first few measurements and is due to the small sample size. As more measurements are added, the filter settles down and provides stable estimates. 

Although they are not shown in the figure, we have also calculated the errors in the individual Euler angles. Because the two initial pure states selected are both sensitive to rotations around the $z$-axis, the errors for the heading angle ($\theta_z$) are lower than for the other two components. This is simply because we have selected initial states in the $z$-basis, and but it would be possible to alternate between states defined using the three principal axes, as we have done with the single qubit case, but it is not strictly necessary in this case. Unlike the single qubit case, all three angles can be estimated from two states constructed in one basis. 
\begin{figure}
    \centering
    \includegraphics[width=0.95\hsize]{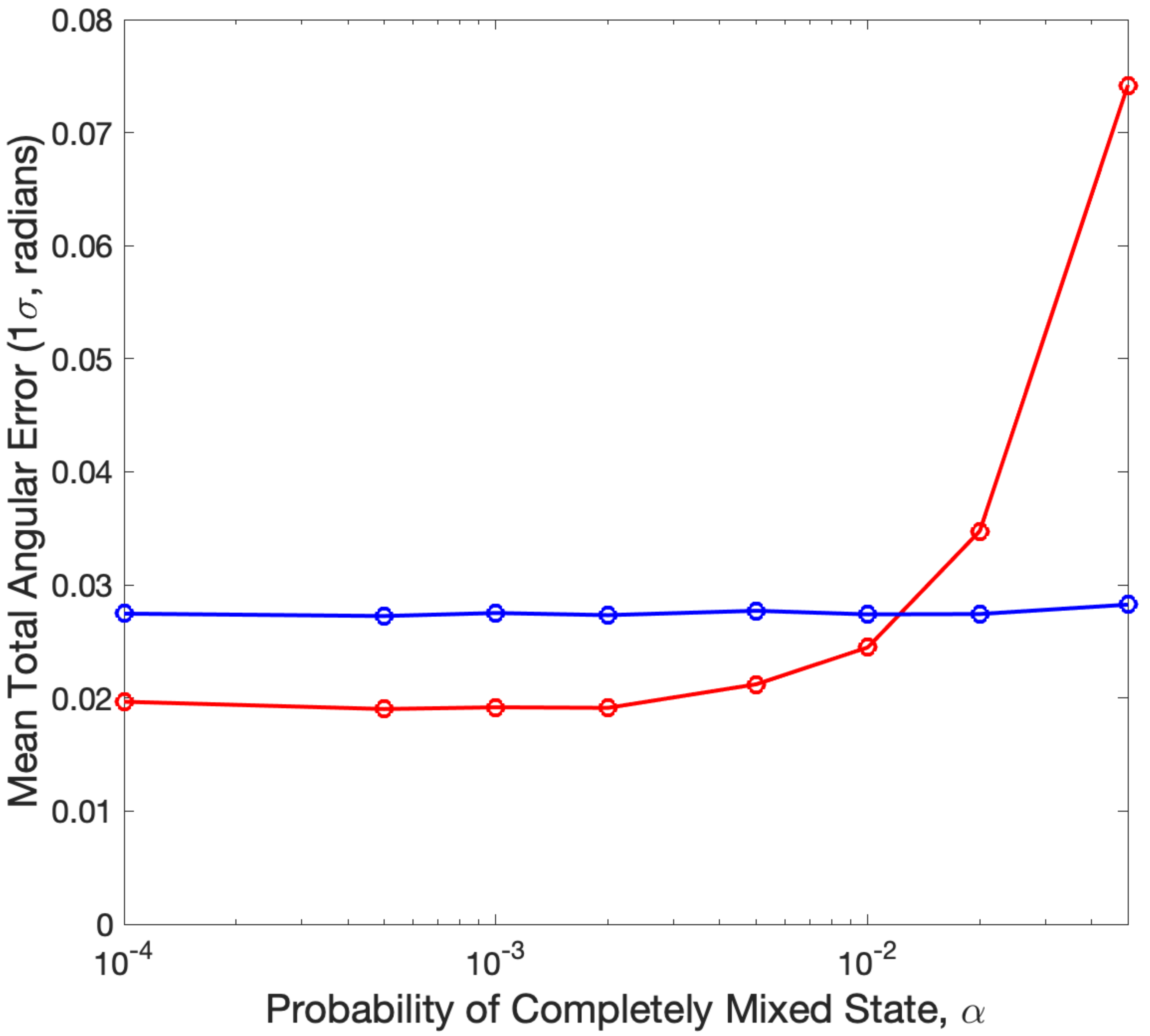}
    \caption{Mean Total Angular Errors (resource count = 8,000) for mixed states as a function of the probability of a completely mixed state, single qubit measurements (blue) and Bell state measurements (red).}
    \label{fig:mixed}
\end{figure}

\section{The effect of mixed states}
\label{sec:sec5}
The results shown in Figure~\ref{fig:errors} assumes that the states are pure initially, and remain pure during the subsequent time period during which the system undergoes a rotation. This is unlikely to be realistic since in any practical system there will be some form of noise that causes dephasing and/or depolarisation. The effect is that the measured state will be mixed rather than pure. To represent a mixed state, we use the density operator representation for the state, $\rho_{Z} = \ket{Z}\bra{Z}$, and we form a mixture by adding a component that is a completely mixed state (i.e., a completely unknown state), which is represented by an identity matrix of the appropriate size, $I_{n\times n}$. So, for the two initial states we are using, we have,
\begin{eqnarray}
    \rho_{\Phi^+} = (1-\alpha)\ket{\Phi^+}\bra{\Phi^+}+\alpha I_{4\times 4}/4 \\
    \rho_{\Phi^-} = (1-\alpha)\ket{\Phi^-}\bra{\Phi^-}+\alpha I_{4\times 4}/4 
\end{eqnarray}
where $\alpha$ is used to measure the effect of incoherent noise. We assume the worst case scenario where noise leads to mixing with a maximally mixed state $I_{4\times 4}/4$. In Figure~\ref{fig:mixed}, we plot the average total error in the Euler angles close to the limit shown in Figure~\ref{fig:errors} for different levels of $\alpha$. We see that the single qubit case is relatively insensitive to the effect of mixed states, but the Bell state case is, as might be expected, quite sensitive to the level of the mixture. There is a significant benefit in using Bell states for $\alpha < 0.005$, but the benefits disappear above about $\alpha \simeq 0.01$. This type of behaviour is not uncommon in quantum metrology, where the benefits associated with the use of entangled states are quite fragile in the presence of noise and uncertainties.

\section{Conclusions}
\label{sec:sec6}
We have presented a Bayesian approach to the estimation of three-dimensional rotations, the Euler angles, using measurements of Bell states formed from spin-$1/2$ systems. The estimation method is based on a particle filter, which is a method taken from Bayesian inference and signal processing. Only three of the Bell states are sensitive to rotations, $\ket{\Phi^\pm}$ and $\ket{\Psi^+}$, and we use the properties of these three states, when constructed in arbitrary bases, to identify the axes to use for the Bell measurements. These axes are selected to maximise the information from the measurements, which we take to be when the probabilities are equal to obtain each of the three states from projective Bell state measurements. We have shown that using the method described to process these Bell state measurements is able to provide a $\sqrt{2}$ improvement in accuracy over the optimal method for individual spin-$1/2$ states. This improvement is in line with what is expected from maximally entangled two-spin-$1/2$ states. We have also considered the effect of noise and dephasing on the performance of the estimation process, using mixed rather than states. And, we find that the benefits from using entangled states decreases more rapidly than for the single spin-$1/2$ states, which is common with quantum metrology protocols based on entangled states. 

\acknowledgments 
The authors would like to thank the UK Quantum Technology Hub Sensors and Metrology for PhD funding related to the work in this paper.

\subsection{Data Availability}
The data that support the findings of this study are available from the corresponding author
upon reasonable request.

\subsection{Conflict of interest}
The authors have no conflicts to disclose.

\bibliography{main}

\end{document}